# Thickness-Dependent and Magnetic-Field-Driven Suppression of Antiferromagnetic Order in Thin $V_5S_8$ Single Crystals

*Will J. Hardy,[†#] Jiangtan Yuan,[‡#] Hua Guo,[‡] Panpan Zhou,[§] Jun Lou,[‡]\* and Douglas Natelson[‡§]\**

[†]Applied Physics Graduate Program, Smalley-Curl Institute, Rice University

Houston, Texas 77005, United States

[‡]Department of Materials Science and NanoEngineering, Rice University

Houston, Texas 77005, United States

[§]Department of Physics and Astronomy, Rice University

Houston, Texas 77005, United States

\*E-mail: jlou@rice.edu

\*E-mail: natelson@rice.edu

**Abstract**

With materials approaching the 2d limit yielding many exciting systems with intriguing physical properties and promising technological functionalities, understanding and engineering magnetic order in nanoscale, layered materials is generating keen interest. One such material is $V_5S_8$, a metal with an antiferromagnetic ground state below the Néel temperature $T_N \sim 32$ K and a prominent spin-flop signature in the magnetoresistance (MR) when $H\|c \sim 4.2$ T. Here we study nanoscale-thickness single crystals of $V_5S_8$, focusing on temperatures close to $T_N$ and the evolution of material properties in response to systematic reduction in crystal thickness. Transport measurements just below $T_N$ reveal magnetic hysteresis that we ascribe to a metamagnetic transition, the first-order magnetic field-driven breakdown of the ordered state. The reduction of crystal thickness to $\sim 10$ nm coincides with systematic changes in the magnetic response: $T_N$ falls, implying that antiferromagnetism is suppressed; and while the spin-flop signature remains, the hysteresis disappears, implying that the metamagnetic transition becomes second order as the thickness approaches the 2d limit. This work demonstrates that single crystals of magnetic materials with nanometer thicknesses are promising systems for future studies of magnetism in reduced dimensionality and quantum phase transitions.

**Keywords:** $V_5S_8$, magnetic layered material, quantum critical point, antiferromagnet, magnetoresistance

The family of two-dimensional materials has grown considerably since the discovery of graphene a decade ago. These 2d materials have demonstrated many exotic and exciting electronic, optical, optoelectronic, mechanical, thermal and catalytic properties.[1-4] Surprisingly missing from this long list are the 2d magnets that could be critical for spintronics applications and fundamental studies of magnetism at the 2d limit. One strategy proposed is to dope 2d materials with transition metals.[5,6] Another possible route is to explore the van Hove singularity often found in 2D materials.[7,8] Layered compounds that possess intrinsic magnetic properties in bulk format such as $CrSiTe_3$ are also proposed to be viable candidates if 2d forms of these materials can be successfully obtained.[9,10] Despite these early efforts, investigation of 2d magnetism is still in its infancy.

$V_5S_8$ (also reported as $VS_{1.64}$ or $V_{1.25}S_2$) has a monoclinic NiAs-type crystal structure with space group C2/m (Figure 1a, b) and an antiferromagnetic ground state, with $T_N \sim 32$ K.[11–13] This material can be relabeled $V_{0.25}VS_2$, to reflect that a fraction of the vanadium atoms are intercalated within the van der Waals gap between $VS_2$ layers (Figure 1a).[14] Early NMR measurements showed that the V sites are not all equivalent[15,16]; neutron diffraction measurements indicate that the $V^I$ sites (the intercalated V atoms, oxidation state $V^{3+}$) have localized 3d electrons with local moments at least $\sim 1.1$ $\mu_B/V^I$, while the $V^{II}$ and $V^{III}$ sites' 3d electrons are itinerant, resulting in metallicity.[11,17–19] Experiments[20] at helium temperatures showed a low-field $H \| c$ positive MR sharply switching to negative MR above $\sim 4.2$ T. While initially modeled as field-induced suppression of AFM order,[18] this sharp feature was identified in bulk samples as the spin-flop transition,[14,18] with the magnetic easy axis oriented[17,20] $\sim 10°$ away from $c$ toward the $a$ axis. Figure S7 shows the magnetic structure of the bulk material and the spin-flopped configuration. Band structure calculations[21] and neutron diffraction[17] estimate a moment $\sim 1.4$ $\mu_B/V^I$. Pulsed high-field magnetization measurements[22] found two anomalies in $M(H)$, the spin-flop at $\sim 4.5$ T and an unknown "distinct hysteresis" at 16.2 – 18.4 T.

We report magnetotransport measurements on thin single crystals revealing MR hysteresis beginning near zero $H$ just below $T_N$, with the hysteresis field scale growing dramatically with decreasing *T*. We surmise that this MR hysteresis is a first-order field-driven transition to a paramagnetic (PM) state. Decreasing thickness down to ~ 10 nm depresses $T_N$ and suppresses the MR hysteresis, while preserving the spin-flop transition. This suggests that reduced dimensionality increases the importance of fluctuations, driving the field-driven transition toward second order and implying the existence of a field-driven quantum phase transition. These investigations show that thin crystals grown by chemical vapor deposition (CVD) are powerful tools for examining magnetism and phase competition in reduced dimensionality.

## Results and Discussion

Thick, shiny, yellow crystals with hexagonal or half hexagonal shapes, expected for the C2/m symmetry of $V_5S_8$,[21] are the most common CVD synthesis products (Fig. 1c shows an optical micrograph of a representative sample). Very thin crystals (as thin as ~ 10 nm) also form, appearing light blue or purple on the growth substrate (Fig. 1d). Figure 1d's inset shows the AFM height profile of the pictured crystal with thickness ~20 nm.

We confirm the structure and composition of the $V_5S_8$ crystals *via* energy dispersive X-ray spectroscopy (EDS) and electron diffraction (ED) (Figure 1e-i). The crystal structure is identified by acquiring selective area electron diffraction (SAED) patterns at various crystal tilts. The derived lattice parameters (a=11.375 Å, b=6.648 Å, c=11.299 Å) are consistent with those of $V_5S_8$. Figures 1j and k are the optical and SEM images of a typical crystal prepared with electrodes.

Temperature-dependent resistivity, $\rho(T)$, measured from 2–300 K(solid lines in Fig. 2, normalized to T = 300 K value) is consistent with metallic conduction and agrees well with prior measurements of a large bulk single crystal of $V_5S_8$,[11] with room temperature resistivity of a few hundred $\mu\Omega$-cm. Upon cooling a broad decrease in resistivity is observed, centered at ~100 K; with further cooling a kink appears, marking $T_N$, whose sharpness and value decrease with decreasing sample thickness. This sharp resistivity drop is attributed to reduction of spin fluctuations upon antiferromagnetic ordering[23]. The resistivity ratio $\varrho(300K)/\varrho(2K)$ falls with decreasing sample thickness, likely due to enhanced interface scattering at small thickness. Plotting d$\varrho$/d$T$ as a function of temperature (Fig. 2, open red circles) allows more accurate determination of the transition temperature, in the range of 22–32 K for different thicknesses, with $T_N$ decreasing monotonically with thickness. In the thinnest samples (~10 nm), an upturn in resistivity is observed below ~ 10 K, which may be a sign of localization. There is no evidence of temperature hysteresis upon subsequent warming.

Small crystal size makes it very difficult to *directly* probe antiferromagnetic order *in situ*. Neutron scattering requires far larger sample mass to produce detectable signals. Combining a technique like x-ray magnetic linear dichroism (XMLD) with photoemission electron microscopy (PEEM) at cryogenic conditions is extremely challenging. Fortunately the distinct spin-flop MR features reported in the prior work on bulk $V_5S_8$ give us access to some of this information. We focused on temperatures just below $T_N$, with $H$ oriented either perpendicular or parallel to the *ab* plane. The MR is defined here as $\Delta\varrho/\varrho = [\varrho(H)-\varrho(0)]/\varrho(0)$. We find sharp, nontrivial MR features appearing only below $T_N$, and in Figure 3 present two families of curve shapes for samples of different thickness (24, 66, and 345 nm thicknesses are the first type; 11, 12, and 13 nm thicknesses are the second).

Two relatively thick, simultaneously grown samples, ~ 66 nm and ~ 345 nm thick, display a strikingly hysteretic and anisotropic series of MR responses with $H$ either perpendicular or parallel to *c*. The 66 nm thick sample's pronounced MR hysteresis appears just below $T_N$ ~ 32 K (Fig. 3a). With $H\|c$, the MR shows the sharp spin-flop downturn at ~ 4 T. Hysteretic MR over a limited $H$ range is observed for temperatures near $T_N$. For example, at T = 30 K, the MR hysteresis loop extends from ~ 5 T to ~ 7 T, with the MR retracing itself outside that field range. The hysteresis quickly shifts to much higher fields as $T$ is reduced. Hysteresis is no longer visible below ~ 20.5 K, having presumably shifted beyond the 14 T limit of our magnet, as suggested by magnetization data at helium temperatures.[17] The sense of the hysteresis is that the resistance is lower on increasing magnitude of field than when sweeping from high field toward zero. In order to check for any effects of the magnetic field ramping rate, selected MR curves were measured using different field sweeping rates (either 25 Oersted/s or 50 Oersted/s), with no discernible differences in the results. Sweep rates are limited to relatively slow speeds to avoid any complications from eddy current heating and the strong temperature dependence of the MR.

When the magnetic field is instead applied in the *ab* plane, the low-field sign of MR remains negative over the whole field range, with no sharp features (Fig. 3b), consistent with the spin-flop interpretation of the sharp MR feature in the $H\|c$ data.[17] Hysteretic MR of comparable magnitude is observed with $H\|ab$, at the *same* field ranges as the $H\|c$ case. This implies that the hysteresis is a first-order field-driven phase transition, driven by magnetic tuning of the relative free energies of the AFM and PM states; alternate explanations involving, *e.g.*, domains within the material would be expected to depend strongly on field direction due to geometric and crystallographic anisotropies. The 345 nm thick sample shows quantitatively similar hysteresis below $T_N$ ~ 32 K, though the signal-to-noise is worse (Fig. S5) because of the thicker sample's very low four-terminal resistance (~ 1 Ω).

A third, considerably thinner sample, 24 nm thick and grown independently from the two thicker crystals, has MR curves with $H\|c$ below $T_N$ ~ 31 K (Fig. 3c) qualitatively similar to those of the 66nm and 345 nm samples, but without discernible hysteresis. Minor $R(H)$ variation likely results from slight temperature drift and/or contact resistance noise, but any true MR hysteresis is too small to detect. In contrast, the $H\|ab$ measurement on the 24 nm sample shows monotonic negative MR and clearly resolved hysteresis (Fig. 3d), very similar to what was observed for the 66 nm sample. This intermediate thickness value demonstrates gradual evolution of the MR properties from thick to thin samples.

For a group of still thinner samples (11, 12, and 13 nm thick samples measured; 12 nm data shown in Fig. 3, and 11 nm in Fig. S3) with $H\|c$, the low temperature MR is fairly flat at small fields, but with increasing field the spin-flop kink begins at ~ 3 T and the MR (a few percent) is negative (see Fig. 3e). For these thin samples $T_N$ is suppressed – for the sample shown (d$\varrho$/d$T$ gives $T_N$ ~ 22 +/- 1 K) the spin-flop kinks are still visible at 23 K and disappear with increasing temperature. At 25 K and above, the MR is more nearly parabolic and without kinks, gradually flattening as the temperature is increased toward 100 K. When $H\|ab$, smooth MR curves without kinks are recorded in the range 20 – 25 K (Fig. 3f). The flat (rather than positive) MR in the low field range for thin samples may be due to a relatively stronger contribution of spin disorder to the zero-field resistivity, weakening the contribution of the spin-flop effect to the MR.

These results demonstrate two features of $V_5S_8$ that were not explained in the prior literature: Field driven first-order breakdown of the AFM state, resulting in MR hysteresis, and thickness dependent suppression of the AFM state as the thickness approaches 10 nm. There are several examples of field-driven metamagnetic transitions involving the breakdown of an AFM state, as in the rare earth containing systems $Nd_{0.5}Sr_{0.5}MnO_3$,[24] $DySb$,[25] and $DyCuSi$ and $HoCuSi$.[26,27] In the case of $Nd_{0.5}Sr_{0.5}MnO_3$, application of a magnetic field up to 12 T at low temperatures causes the insulating AFM state to break down to a metallic FM state, accompanied by a drop in resistivity of more than five orders of magnitude.[24] Alternatively, a metamagnetic transition can occur between AFM and PM order, as in the case of CeAuSb$_2$, where hysteretic MR measurements suggest a first order transition. This material is therefore thought to be very near to a quantum critical point (QCP).[28] While it is conceivable that the hysteresis is an indicator of some other ordered state (*e.g.*, charge density wave), this is unlikely given the lack of any previous evidence for competing order or magnetoelectric effects in this material.

The observed thickness dependence could be intrinsic or extrinsic. Confinement can affect the itinerant carriers, as seen in the thickness-dependent $\rho(T)$, potentially impacting exchange processes. The effects of uncompensated surface spins may also be at play,[29] due to a possible lack of perfect AFM order at the crystal surfaces, which would be more important in crystals of reduced thickness. Thinner crystals are more strained due to their elastic interactions with the underlying substrate. Stoichiometry could also be a concern, though there is no evidence of systematic variation of V:S ratio with thickness.

Combining resistivity and MR information, we construct a tentative phase diagram using temperature and critical field values (Fig. 4). Each filled data point is taken from an MR isotherm such that the point's *x*-axis position represents the central value of a MR hysteresis loop at a given temperature (scaled by $T_N$), and the horizontal bar represents the hysteresis loop width. Data for the 345 nm and 66 nm samples were extracted from *H*//*c* measurements, while for the 24 nm sample, which displayed no apparent *H*//*c* hysteresis, the data points were taken from the hysteretic *H*||*ab* MR curves. Since our maximum field strength was 14 T, data points at the lower end of the temperature range have horizontal bars limited by the accessible field. These phase assignments are based on the appearance of the previously reported spin-flop kink in the MR, with PM as the logical competing phase based on the onset of the metamagnetic transition at $T_N$, as there is no direct technique available to probe *in situ* AFM order in these very small samples.

The plot's useful range can extend to lower temperatures by including the single high-field data point at T = 4.2 K from *M(H)* measurements on a bulk polycrystalline sample,[22] presuming that this *M(H)* hysteresis results from the same AFM breakdown phenomenon as our MR hysteresis close to $T_N$. Fig. 4 implies a AFM/PM quantum phase transition at zero temperature and finite magnetic field value $H_C(0)$, which can be estimated by fitting: $H_C = H_C(0) \times (1-T/T_N)^\alpha$. The fit yields an estimate of $H_C(0)$= 18.7 T and an exponent $\alpha$ = 0.345, which deviates from the value of ½ previously seen in other antiferromagnets.[30,31]

## Conclusions

The field-driven breakdown of the AFM state implies the existence of a quantum phase transition at accessible fields in this material as $T \rightarrow 0$. These studies also suggest a suppression of antiferromagnetism with reducing thickness, although even very thin crystals (~ 10 nm) maintain some AFM signatures (spin flop kinks in the *H*||*c* MR). Given the suppression of $T_N$ in the thinnest crystals, it is unlikely that the metamagnetic transition is absent in those samples, even without its clearest evidence (the hysteretic MR). The disappearance of MR hysteresis as thickness is reduced suggests that the metamagnetic transition changes from first order to second order. This would imply possible quantum critical phenomena at appropriately large field scales, though these have not yet been examined. Further experiments at higher fields and lower temperatures (*e.g.*, noise/resistance fluctuations; magneto-optic response; XMLD with PEEM) are called for and beyond the scope of the present work. While very challenging to implement, XMLD with PEEM *in situ* in the magnetic field regime of the observed metamagnetic transition could potentially examine the hysteretic state (*e.g.*, domains and phase separation) and its evolution with thickness. The rich phenomenology seen in the $V_5S_8$ system toward the 2D limit strongly suggests that CVD growth of other magnetically ordered, layered materials will enable insights into the competition between electronically and magnetically ordered phases in reduced dimensionality.

**Methods**

Single crystals were grown by CVD on 300 nm $SiO_2$/Si wafers, with lateral crystal dimensions ranging from a few microns to a few tens of microns, and thicknesses from ~10 nm to ~1 um. Vanadium trichloride ($VCl_3$) was used as a precursor (Sigma-Aldrich, 97%, powder). An $Al_2O_3$ crucible with around 0.1 g of $VCl_3$ was placed in a quartz tube with diameter of 1 inch at the center of the heating zone of the furnace, and a $SiO_2$/Si substrate was placed on top of the crucible with oxide surface facing down to collect the final product. Sulfur (1g) was put in the upstream of the tube (Sigma-Aldrich, reagent grade). A mixture of $N_2$ with 10% of $H_2$ (99.999%) was used as carrier gas at ambient pressure. The center of the furnace was gradually heated from room temperature to 750 °C in 25 min at a rate of ≈20 °C /min. After 10 minutes at this temperature, the carrier gas was changed to pure $N_2$ (99.999%) and the furnace was naturally cooled down to room temperature. Thin crystals are usually formed near the edge of growth substrate.

Metal contacts were defined on the as-grown crystals by a standard procedure of e-beam lithography, e-beam evaporation of Au with a V adhesion layer, and liftoff. For relatively thick samples, an additional step to deposit extra Au (40 nm) by dc sputtering was included between the e-beam evaporation and liftoff steps in order to ensure that the electrodes had continuous metal coating over the crystal side walls. The completed devices were characterized by optical microscopy, scanning electron microscopy (SEM) and atomic force microscopy (AFM), as shown in Fig.1. Transport measurements were performed using four-probe low frequency lock-in methods in either a Quantum Design Physical Property Measurement System (QD PPMS) equipped with a 9 T superconducting magnet or a QD Dynacool PPMS equipped with a 14 T superconducting magnet. For some measurements, a QD horizontal rotator apparatus was employed to change the relative direction of the applied magnetic field with respect to the sample axes.

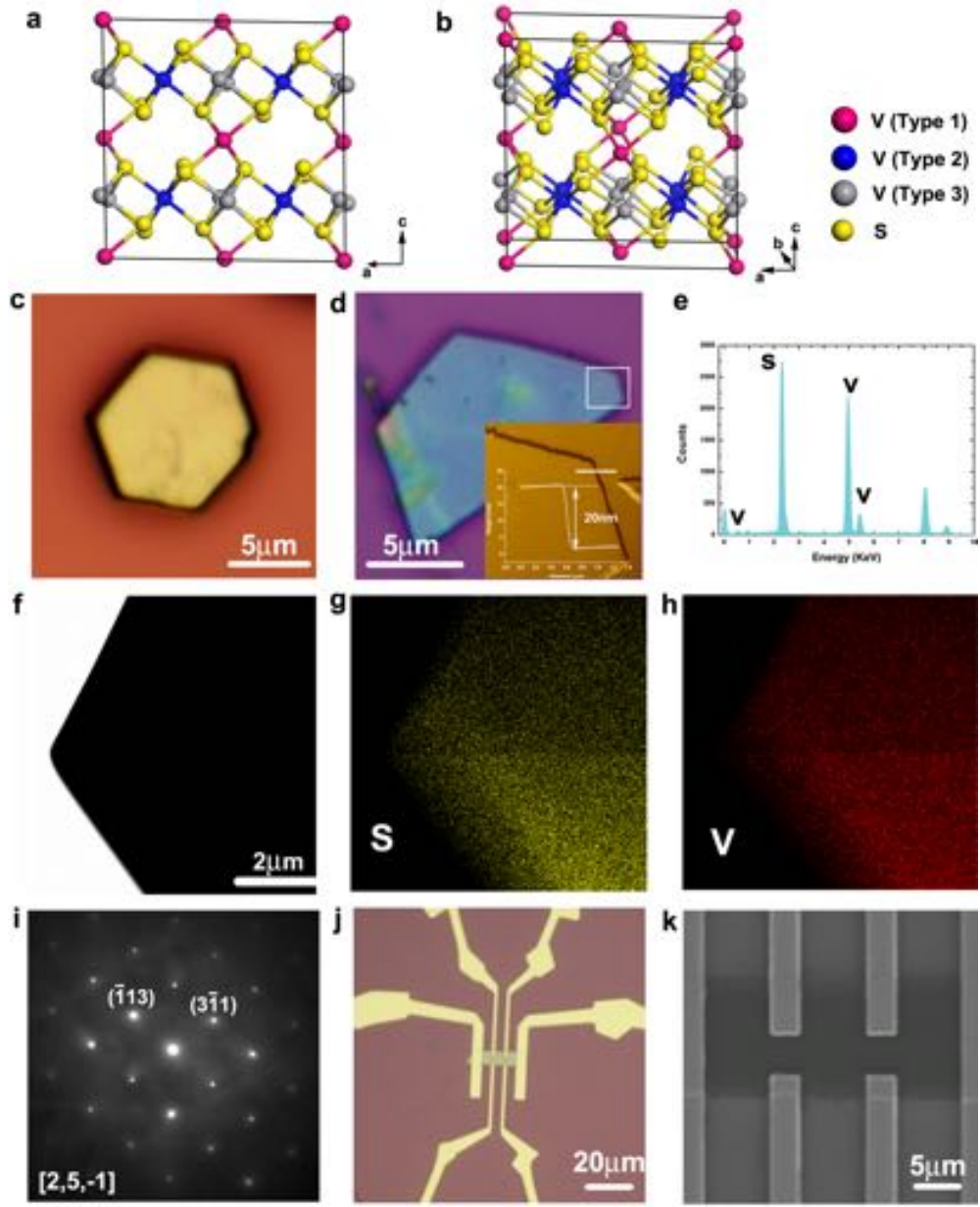

**Figure 1. Structure and characterization of thin $V_5S_8$ crystals and devices**: a, b, the crystallographic unit cell of $V_5S_8$. Different colors are used to distinguish three types of vanadium atoms. From the side view along b axis in a, $V_5S_8$ can be considered as intercalating vanadium atoms within the van der Waals gap between layers of $VS_2$. Optical images of (c) a thick and (d) a thin crystal. Inset of (d) is the AFM height profile which shows the thin crystal has a thickness of about 20nm.. (e) EDS spectrum of crystal shown in (f) the brightfield TEM

image. EDS peaks for S and V are labeled, while other visible peaks are due to the Cu TEM grid. (g,h) EDS maps of (g) sulfur and (h) vanadium for the crystal in (e), showing a uniform distribution of both elements. (i) SAED pattern of the same crystal along the [2, 5, -1] zone axis, with lattice parameters consistent with $V_5S_8$. (j) Optical and (k) SEM images of a representative device used for electronic transport measurements (66 nm thick).

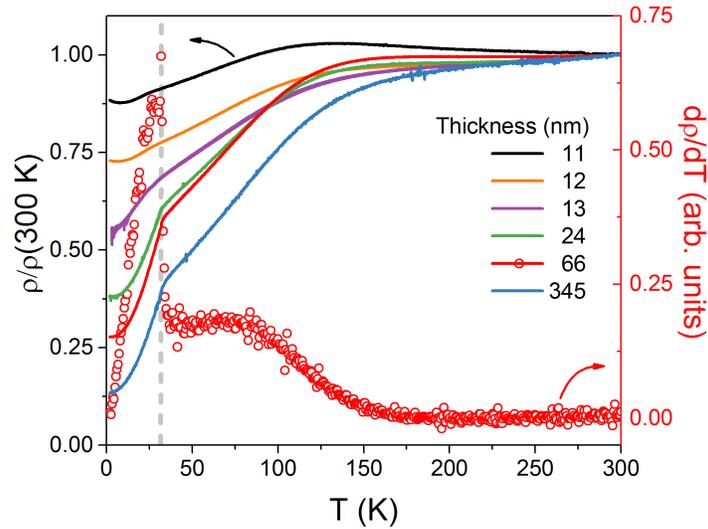

**Figure 2. Resistivity vs. *T*:** Resistivity normalized to T = 300 K value for several crystal thicknesses (solid lines, left axis) and calculated derivative $d\rho/dT$ (open red circles, right axis) of a 66 nm thick crystal as a function of temperature . The sharp kink in resistivity and corresponding peak in $d\rho/dT$ at ~ 32 K mark $T_N$ for the 66 nm sample.

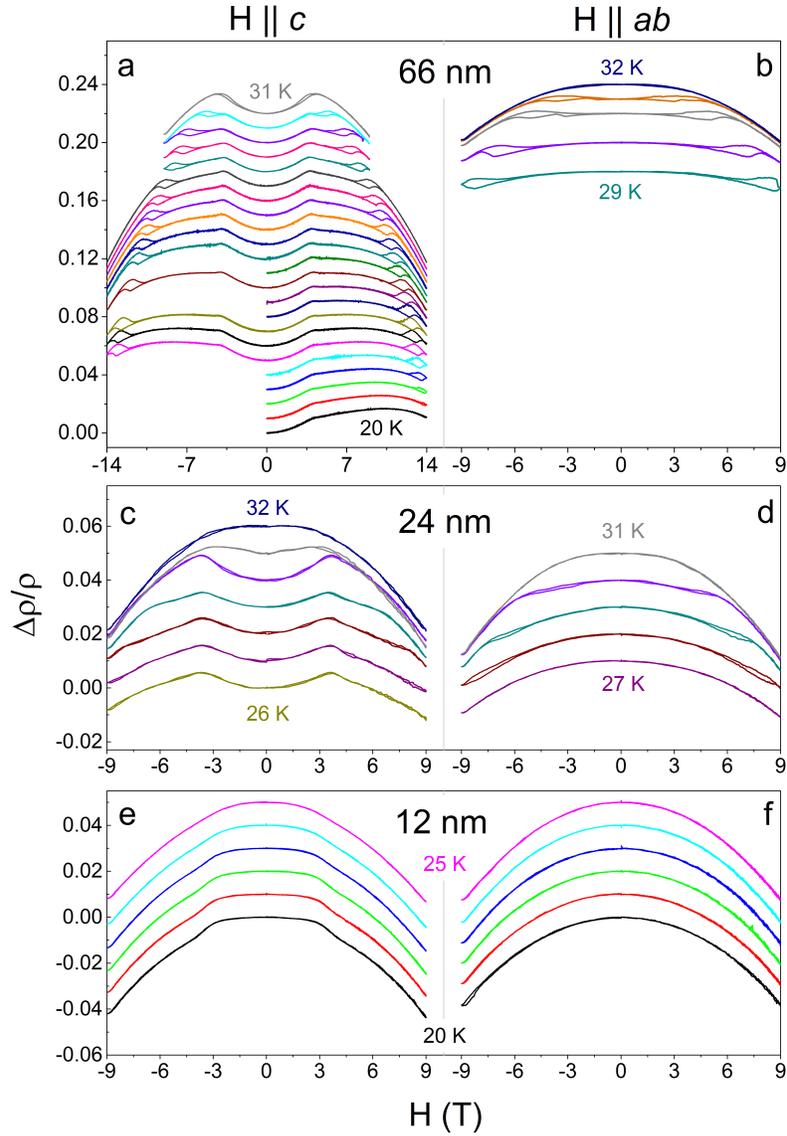

**Figure 3. Magnetoresistance at various crystal thicknesses**: Magnetoresistance curves at selected temperatures for three different crystal thicknesses, with *H*||*c* (left column) or *H*||*ab* (right column). Curves at different temperatures are offset for clarity, with the offset in the left and right column plots matching for curves at the same temperature. Note that for panel (a) the field scale is 14 T, rather than the 9 T scale of the other panels. (a,b) MR measurements for the *t* = 66 nm sample focusing on the temperature range 20–32 K (curves offset by 0.01 per 0.5 K) to demonstrate the evolution of hysteresis, with hysteresis loops shifting to higher fields with decreasing temperature. The peaks near 4 T in the *H*||*c* curves are due to the spin flop transition, whereas the *H*||*ab* curves are negative over the entire field range. (c,d) MR curves for the $t$ = 24 nm sample from 26–32 K (curves offset by 0.01 per 1 K above 26 K). The *H*||*c* curves show no hysteresis but otherwise have qualitatively similar features to those observed for thicker samples. Hysteresis appears when *H*||*ab*. (e,f) MR curves for the $t$ = 12 nm sample in the temperature range 20–25 K (curves offset by 0.01 per 1 K above 20 K). With $H||c$, a smooth and gradual development of kinks is visible near H ~ 3 T as the temperature is decreased, but no hysteresis is observed. With H || *ab*, the curves are smooth and without kinks or discernible hysteresis.

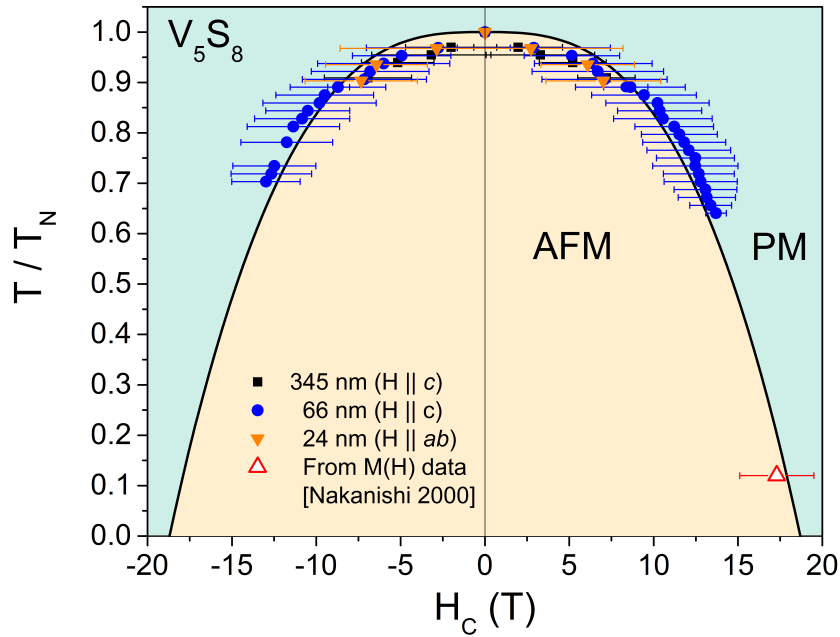

**Figure 4. Suggested magnetic phase diagram for V$_5$S$_8$:** Temperature-magnetic field phase diagram of V$_5$S$_8$. Filled symbols represent the central values H$_C$ of MR hysteresis loops extracted from the data sets of three devices (H ⊥ plane for 345 nm and 66 nm samples, and H || plane for 24 nm sample), and corresponding horizontal bars represent the width of those hysteresis loops.

The open red triangle represents one data point from high field magnetization measurements[17] at T = 4.2 K. The solid line is a fit inspired by a functional form used previously in the field-driven breakdown of AFM order in the CeIn$_3$ system.[30,31]

## Supporting Information.

Temperature dependence of resistivity for additional samples, additional MR and Hall data, morphologies of additional V$_5$S$_8$ crystals. This material is available free of charge *via* the Internet at http://pubs.acs.org.

## Corresponding Author
*E-mail: jlou@rice.edu

*E-mail: natelson@rice.edu

## Author Contributions

#W.J.H. and J.Y. contributed equally to this work. W.J.H. and P.Z. fabricated devices and performed transport measurements and SEM characterization. J.Y. synthesized the crystals. W.J.H., J. Y. and P.Z. performed AFM characterization. J.Y. and H.G. characterized the crystal structure with TEM. J.L. and D.N. actively guided the study. W.J.H., J.Y., J.L., and D.N. wrote the manuscript. All authors discussed the results and edited the manuscript.

## Acknowledgements

W.J.H., P.Z., and D.N. gratefully acknowledge support from the US DOE Office of Science/Basic Energy Sciences award DE-FG02-06ER46337. J.Y., H.G. and J.L. gratefully acknowledge support from the AFOSR (grant FA9550-14-1-0268) and the Welch Foundation (Grant C-1716). The authors want to thank X. Zou for his help in building the crystal structure models, as well as E. Morosan and B. Rai for their help with high-field MR measurements.

## Competing Financial Interests
The authors declare no competing financial interest.

## Table of Contents Graphic and Synopsis

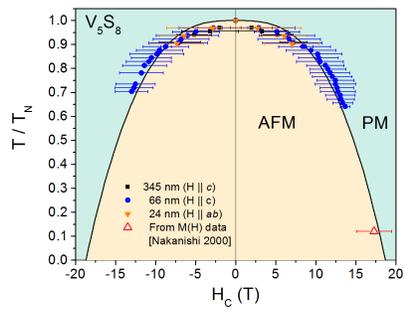

Systematic tuning of thickness of a magnetic layered material ($V_5S_8$) toward the 2d limit reveals a field-driven magnetic transition, hints of a quantum critical point, and evolution of the stability of the magnetic phase.

**Supporting Information for:**

# Thickness-Dependent and Magnetic-Field-Driven Suppression of Antiferromagnetic Order in Thin V$_5$S$_8$ Single Crystals

*Will J. Hardy,[†#] Jiangtan Yuan,[‡#] Hua Guo,[‡] Panpan Zhou,[§] Jun Lou,[‡]\* and Douglas Natelson[‡§]\**

[†]Applied Physics Graduate Program, Smalley-Curl Institute, Rice University

Houston, Texas 77005, United States

[‡]Department of Materials Science and NanoEngineering, Rice University

Houston, Texas 77005, United States

[§]Department of Physics and Astronomy, Rice University

Houston, Texas 77005, United States

## I. Temperature dependence of resistivity for additional samples

For comparison with the temperature dependent resistivity data for the 66 nm thick sample, additional plots are shown in Fig. S1 of the resistivity and its derivative d$\rho$/d$T$ as a function of temperature for the 11, 12, 24, and 345 nm thick samples. Vertical gray dashed lines mark the position of $T_N$ at the peak of d$\rho$/d$T$. The 11 and 12 nm thick samples have a subtle resistivity kink at the transition temperature, whereas the corresponding kink for the thicker samples is sharper and more pronounced. The 11 nm sample shows a negative slope of resistivity at high temperatures, possibly due to contact issues, while the data below ~ 150 K looks very similar to the curve for the 12 nm thick sample. The resistivity curves for both 11 and 12 nm thick samples show an upturn in resistivity at temperatures below ~ 10 K.

The overall variation in magnitude of resistivity among the samples may be due to slight variations in crystal stoichiometry, as well as estimation of the sample dimensions from SEM and AFM images of the hexagon shaped samples, the geometry of which makes proper estimation of lateral dimensions somewhat challenging. We note that the fractional change of resistivity relative to its value at $T$ = 300 K is larger for thicker samples, which may be either an indication of higher quality of thick crystals or of other thickness dependent effects (e.g., surface scattering).

In Fig. S2, we plot the thickness dependence of the Néel temperature $T_N$ as determined from the position of the peak of d$\rho$/d$T$ vs. $T$. A systematic reduction of the transition temperature from its large-thickness saturation value is visible as the thickness is decreased (~ 32 K for the 66 nm and 345 nm thick samples; ~ 31 K for the 24 nm sample; ~ 22 K for the 11 and 12 nm thick samples). An estimated error of ~ ±1 K in determining $T_N$ results from both the width of the transition as well as noise in the calculated d$\rho$/d$T$ vs. $T$ curves.

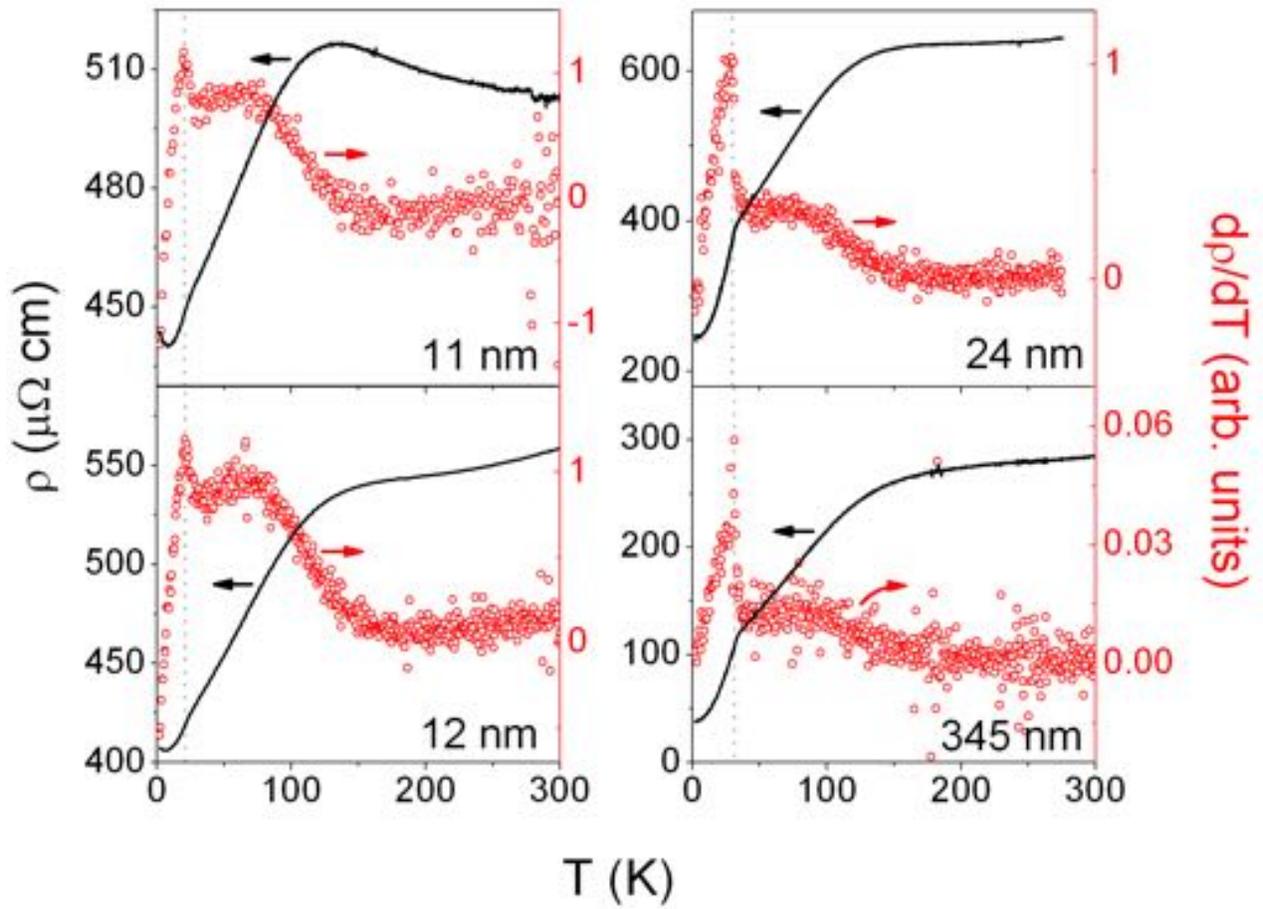

**Fig. S1:** (Left axis, black solid lines) Resistivity and (right axis, open red circles) calculated derivative dρ/dT as a function of temperature for the 11, 12, 24, and 345 nm thick samples. The dashed gray lines mark $T_N$.

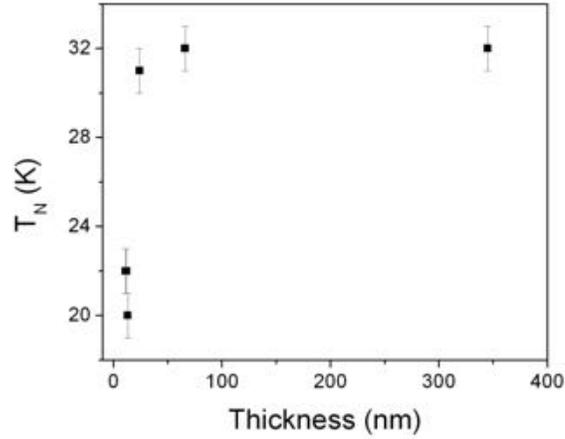

**Fig. S2**: Crystal thickness dependence of $T_N$. A saturation value of ~ 32 K is observed for the thickest samples.

## II. Additional MR and Hall data

For comparison with the measurements presented in the main text, here we provide additional data sets of MR as well as Hall resistivity for the 11, 66, and 345 nm thick samples. The left panel of Fig. S3 shows H||c MR measurements for the 11 nm sample over the wide temperature range 2 – 100 K. Similar to the data for the 12 nm sample, kinks near ~ ±3 T are present at T = 20 K and below, but no hysteresis is observed. Corresponding Hall resistivity measurements are shown in the right panel of Fig. S3. The Hall curves are linear at temperatures above $T_N$ but superlinear at T = 20 K and below, with pronounced kinks near ~ ±3 T and no detectable hysteresis. The raw data were antisymmetrized (to remove the $\rho_{xx}$ contribution that was present due to non-ideal sample geometry and imperfect electrode alignment) using the relation: $\rho_{xy} = [\rho_{xy}(H) - \rho_{xy}(-H)]/2$.

Fig. S4 shows H||c MR isotherms for the 66 nm sample, augmenting the data presented in the main text with measurements over the expanded temperature range 2 – 150 K The data are

shown here without offsets for different temperatures in order to emphasize the nonmonotonic change of the curve shape over the full temperature range. Due to the coarse temperature increment, hysteresis is only observed at T = 30 K. The MR sign is negative at 40 K and above, becoming flat at 100 K and higher. The right panel of Fig. S4 shows the corresponding Hall resistivity data, which has a negative slope at all measured temperatures, consistent with majority *n*-type carriers. The slope decreases with increasing temperature, suggesting a corresponding increase of carrier density with increasing $T$. A slightly superlinear response is observed at fields above ~ ±3 T when the temperature is below $T_N$.

Fig. S5 (left panel) shows H||c MR isotherms for the 345 nm thick sample in the temperature range 28–40 K, confirming the nature of the MR hysteresis observed in the 66 nm thick sample. The onset of hysteresis is observed at ~32 K and disappears as the temperature decreases below ~ 30 K due to the limited 9T range of the magnetic field. The right panel of Fig. S5 shows the corresponding Hall resistivity curves, with a superlinear trend visible at fields above ~ ±3 T when the temperature is below $T_N$, becoming more pronounced as the temperature decreases.

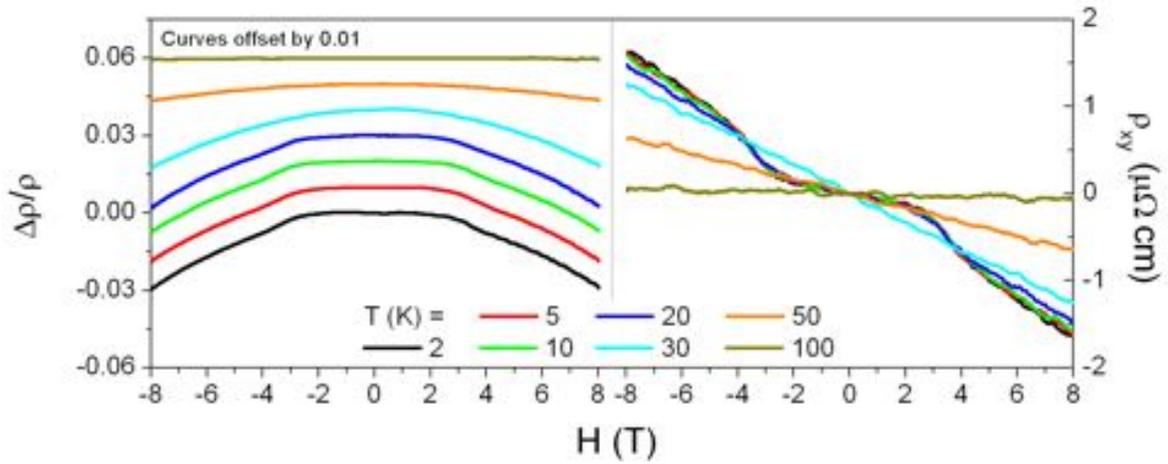

**Fig. S3:** (Left) H||c MR isotherms for the 11 nm thick sample in the temperature range 2 – 100 K. Curves are offset by 0.01 for clarity. (Right) Hall resistivity data for the same sample.

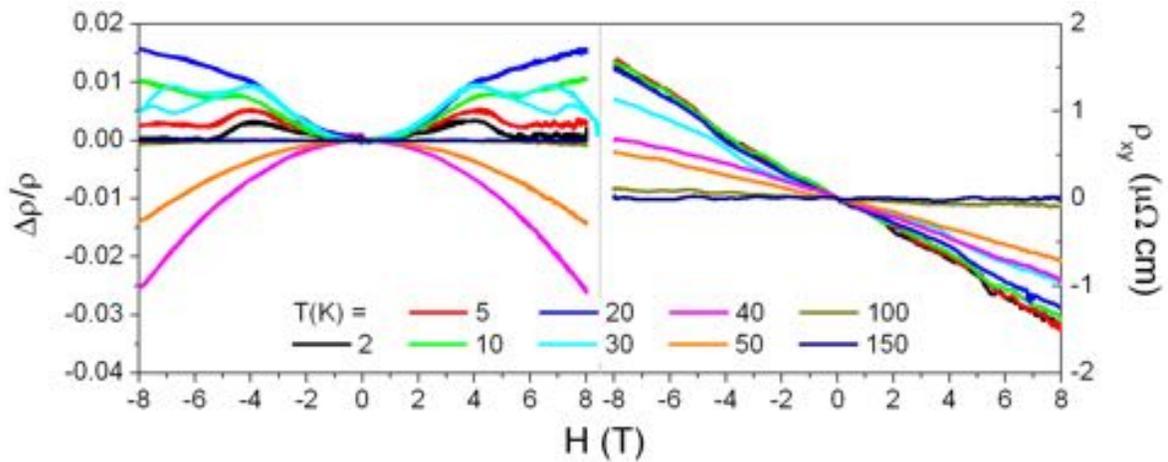

**Fig. S4:** (Left) H||c MR isotherms for the 66 nm thick sample at selected temperatures in the range 2 – 150 K. (Right) Hall resistivity data for the same sample at corresponding temperatures.

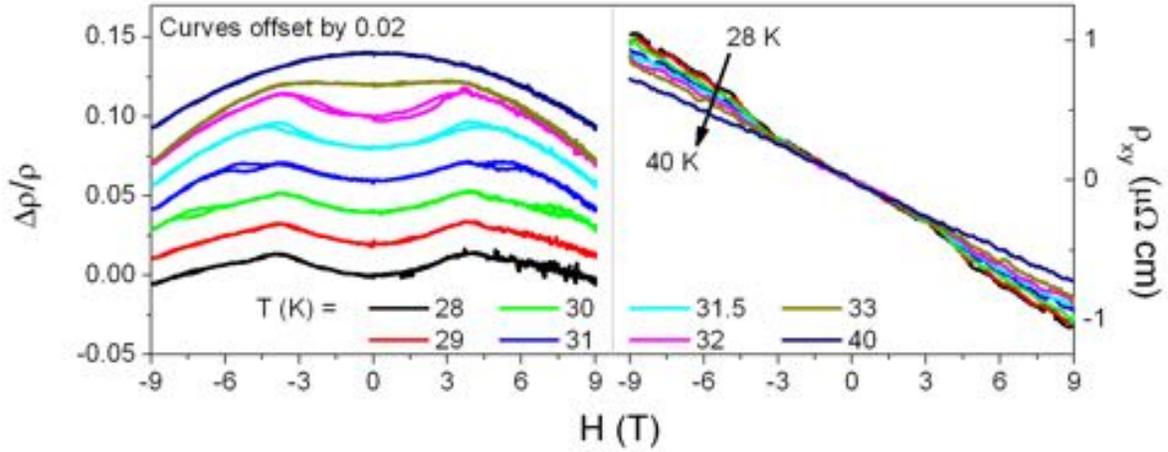

**Fig. S5:** (Left) H∥c MR isotherms for the 345 nm thick sample in the temperature range near $T_N$ from 28–40 K. Curves are offset by 0.02 for clarity. (Right) Hall resistivity curves for the same sample at corresponding temperatures.

### III. Morphologies of additional $V_5S_8$ crystals.

Fig. S6 shows three additional optical micrographs as examples of the hexagonal motif of the CVD-grown $V_5S_8$ thin crystals. The faceted character of the crystals indicates likely single-crystal growth.

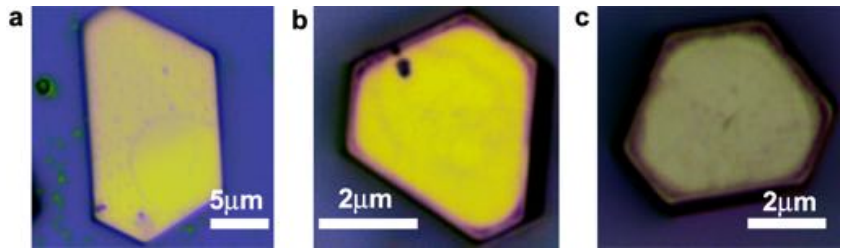

**Fig. S6:** Additional, representative crystals grown along with the ones measured and described in the main text.

## IV. Magnetic structure of $V_5S_8$

Fig. S7 shows the magnetic structure of bulk $V_5S_8$ in the AFM state[1,2]. The magnetic moments are carried by the type-I vanadium atoms. The magnetic easy axis is approximately 10° toward the *a* axis from the *c* axis in the *a-c* plane. When an external magnetic field $H$ greater than ~ 4.2 T (in the bulk) is applied along *c*, the system undergoes a spin-flop transition into the configuration shown in Fig. S7(b).

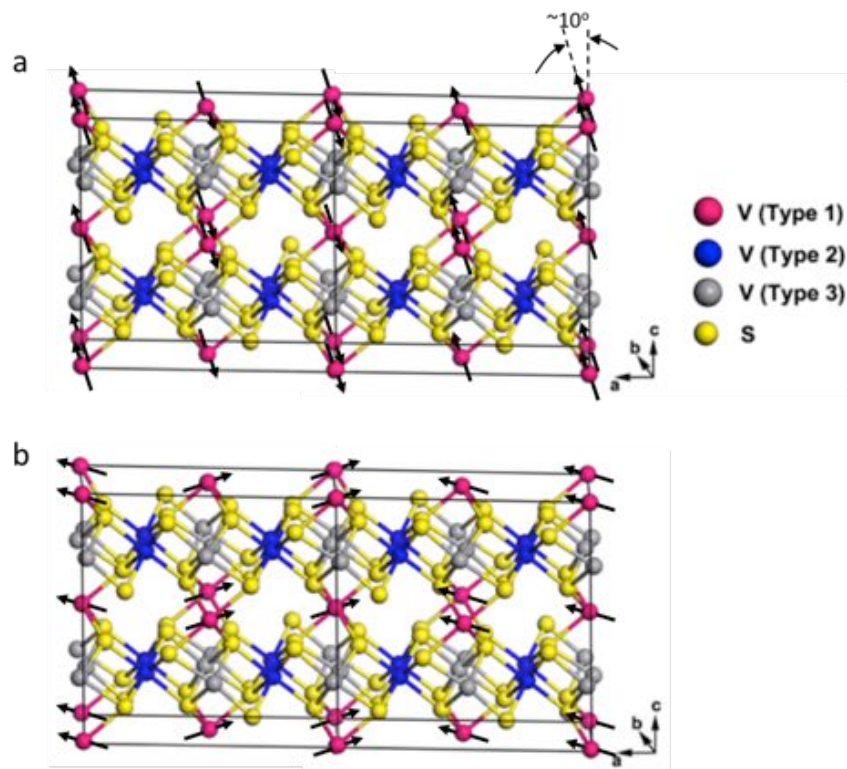

**Fig. S7**. Magnetic structure of $V_5S_8$. (a) The AFM ordered structure in the ordinary Neel state. (b) The conjectured magnetic structure in the spin-flopped state when a large external magnetic field has been applied along the *c* axis.